\documentclass{PoS}
\usepackage[utf8]{inputenc}
\usepackage[T1]{fontenc}
\usepackage{graphicx}
\usepackage{subcaption}
\usepackage{slashed}
\usepackage{dsfont}

\title{
\begin{flushright}\begin{small}
 {\it CP3-Origins-2018-041 DNRF90\\} \vspace{1.0cm}
  \end{small}
\end{flushright}
Taylor expansion and the Cauchy Residue Theorem for finite-density QCD
} 

\ShortTitle{Taylor expansion and the Cauchy Residue Theorem}

\author{Philippe de Forcrand\\
        Institute for Theoretical Physics, ETH, CH-8093 Zürich, Switzerland \& 
        CERN, Theory Department, CH-1211 Geneva, Switzerland\\
        E-mail: \email{forcrand@phys.ethz.ch}}

\author{\speaker{Benjamin Jäger}\\
        CP3-Origins \& Danish IAS, Department of Mathematics and
Computer Science, University of Southern Denmark, Campusvej 55, 5230 Odense M,
Denmark\\
E-mail: \email{jaeger@cp3.sdu.dk}}

\abstract{
We present an update on our efforts to determine the Taylor coefficients 
of the $\mu/T$ expansion of the pressure for finite-density QCD. 
Here, we explore alternatives based on the Cauchy Residue Theorem, 
which allows us to use a discretized contour to determine the desired spectral
moments occurring in the Taylor expansion of QCD at zero chemical potential. 
}

\FullConference{
The 36th Annual International Symposium on Lattice Field Theory - LATTICE2018\\
22-28 July, 2018\\
Michigan State University, East Lansing, Michigan, USA.
}

\begin{document}

\section{Introduction}

Information on the QCD phase diagram can be obtained by expanding the pressure $P(T,\mu)$ in a Taylor series 
in terms of the chemical potential, namely~\cite{Ejiri:2003dc,Allton:2005gk}
\begin{equation}
\frac{P(T,\mu)}{T^4} \equiv \frac{\mathrm{log} Z(T,\mu)}{V\, T^3} = \frac{P(T,\mu=0)}{T^4} + \sum_{k=1} c_{2k}(T) \left(\frac{\mu}{T}\right)^{2k}.
\label{eq.intro.taylor}
\end{equation}  

The Taylor coefficients $c_{2k}(T)$ can be expressed as expectation values of traces of operators evaluated at {\em vanishing}
chemical potential, rendering the approach sign-problem free.
However, the number of relevant terms grows quickly, so that the Taylor coefficients above the eighth order have not been estimated reliably yet. 
An important simplification has been proposed by Gavai and Sharma~\cite{Gavai:2014lia,Gavai:2015ywa},  
who have suggested to use a linear chemical potential instead of the "standard" exponential definition~\cite{Hasenfratz:1983ba}. Then, all operators
$\frac{\partial^n \slashed{D}}{\partial \mu^n}, n>1$ vanish. The remaining terms are
all of the form
\begin{equation}
    \mathrm{Tr} \left[ \left( \slashed{D}^{-1} \frac{\partial \slashed{D}}{\partial \mu}\right)^{k} \right].
    \label{eq.intro.tr}
\end{equation}
The divergences originating from the linear chemical potential can be removed by comparing to the free theory. For Staggered quarks, the Dirac operator $\slashed{D}$ is anti-Hermitian, so that all its eigenvalues are imaginary. The derivative with respect to chemical potential $\frac{\partial \slashed{D}}{\partial \mu}$ is a Hermitian matrix. Unfortunately, the product of both has, in general, complex eigenvalues. 
\begin{figure}
\centering
\begin{subfigure}{.5\textwidth}
  \centering
  \includegraphics[width=.98\linewidth]{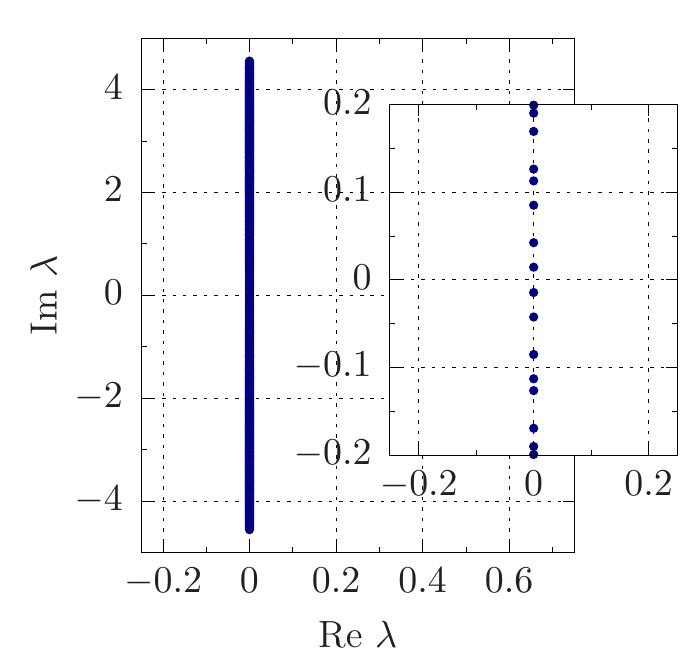}
\end{subfigure}%
\begin{subfigure}{.5\textwidth}
  \centering
  \includegraphics[width=.98\linewidth]{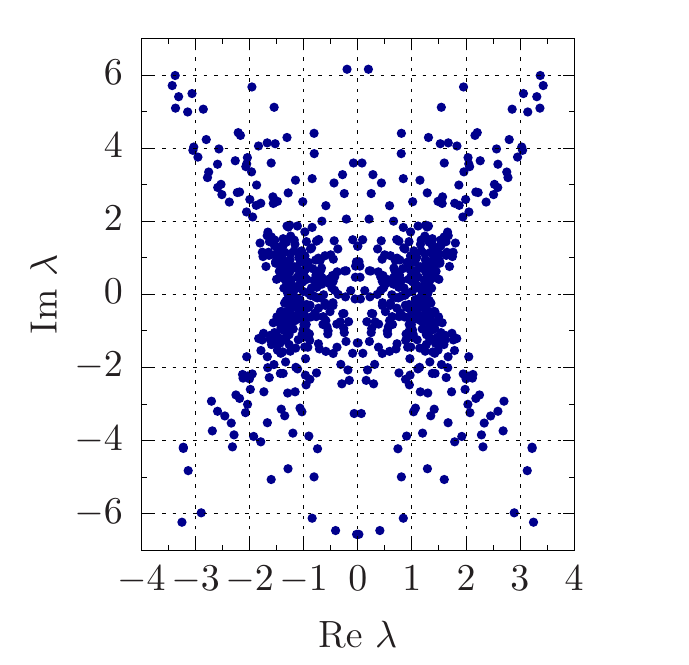}
\end{subfigure}
\caption{Spectrum of the Dirac operator $\slashed{D}$ (left) and $\slashed{D}\, \big(\frac{\partial \slashed{D}}{\partial \mu}\big)^{-1}$ (right).}
\label{fig.spectra}
\end{figure} 
Figure~\ref{fig.spectra} shows the spectrum of $\slashed{D}$ and $\slashed{D}\, \big(\frac{\partial \slashed{D}}{\partial \mu}\big)^{-1}$. 
Fortunately, the trace in equation~(\ref{eq.intro.tr}) is either purely real or imaginary, depending on the order $k$, since for $k=1$ it has the
form $\mathrm{Tr}\left[ A  B \right]$, with $A$ anti-hermitian and $B$ hermitian, and
\begin{equation}
\mathrm{Tr}\left[ A  B \right] = \frac{1}{2} \left( \mathrm{Tr} \left[ A  B \right] + \mathrm{Tr}
\left[ B  A \right] \right) = \frac{1}{2} \left( \mathrm{Tr} \left[ A  B - (A B)^\dagger \right] \right). 
\label{eq.intro.prod}
\end{equation}
Thus, terms of even order are real, as appropriate for equation (\ref{eq.intro.taylor}).
Here, we use four flavours of Staggered fermions and a small $4^4$ lattice size, with a gauge coupling $\beta=5.045$ and a bare quark mass $a m_q=0.07$. For reference, we compute the full spectrum using Mathematica~\cite{Mathematica}. In practice, it turns out to be more efficient to consider
the inverse operator, since both $\slashed{D}$ and $\big(\frac{\partial \slashed{D}}{\partial \mu}\big)^{-1}$ are space-wise sparse:
\begin{equation}
\mathrm{Tr} \left[ \left( \slashed{D}^{-1} \frac{\partial \slashed{D}}{\partial \mu}\right)^k \right] = \mathrm{Tr} \left[ \left( \slashed{D} \Big( \frac{\partial \slashed{D}}{\partial \mu}\Big)^{-1} \right)^{-k}
\right] 
\label{eq.moments}
\end{equation}

\section{Refinement}

The Cauchy Residue Theorem can be used to relate the eigenvalues of a complex matrix to a contour integral. 
For instance, the number $n(\Gamma)$ of eigenvalues inside a closed circle $\Gamma$ of radius $r$ centered at the origin, is given by
\begin{equation}
n(\Gamma) \approx \frac{1}{n_I} \sum \limits_{j=1}^{n_I} \, r\,  \mathrm{e}^{2 \pi \mathrm{i} j / n_I}\, \mathrm{Tr} \left[ \left(z_j \mathds{1} - M\right)^{-1}\right],  \label{eq.refi.eig}
\end{equation}
where the circle has been discretized in $n_I$ points $z_j = r \mathrm{e}^{2 \pi \mathrm{i} j / n_I}$, 
and the integral has been approximated using the trapezoidal rule. 
Other contours can be treated in a similar way. The trace can be estimated using standard Gaussian or $Z_2$ noise vectors. 
In principle, the trace can be obtained using a shifted solver, since the inverse matrix of 
\begin{equation}
M =  \slashed{D}\,\, \left( \frac{\partial \slashed{D}}{\partial \mu}\right)^{-1}
\end{equation}
is just shifted by the quadrature points $z_j$ of the integration. A simple refinement strategy can be devised by starting with a large square domain (covering all eigenvalues), which can be divided into 4 smaller squares if $n(\Gamma) \neq 0$. This division can be repeated until eigenvalues are localised to sufficient precision. 
Then, all moments $k$ in equation~(\ref{eq.intro.tr}) can be obtained, since the whole spectrum is known to good accuracy.
\begin{figure}
\centering
\begin{subfigure}{.5\textwidth}
  \centering
  \includegraphics[width=.98\linewidth]{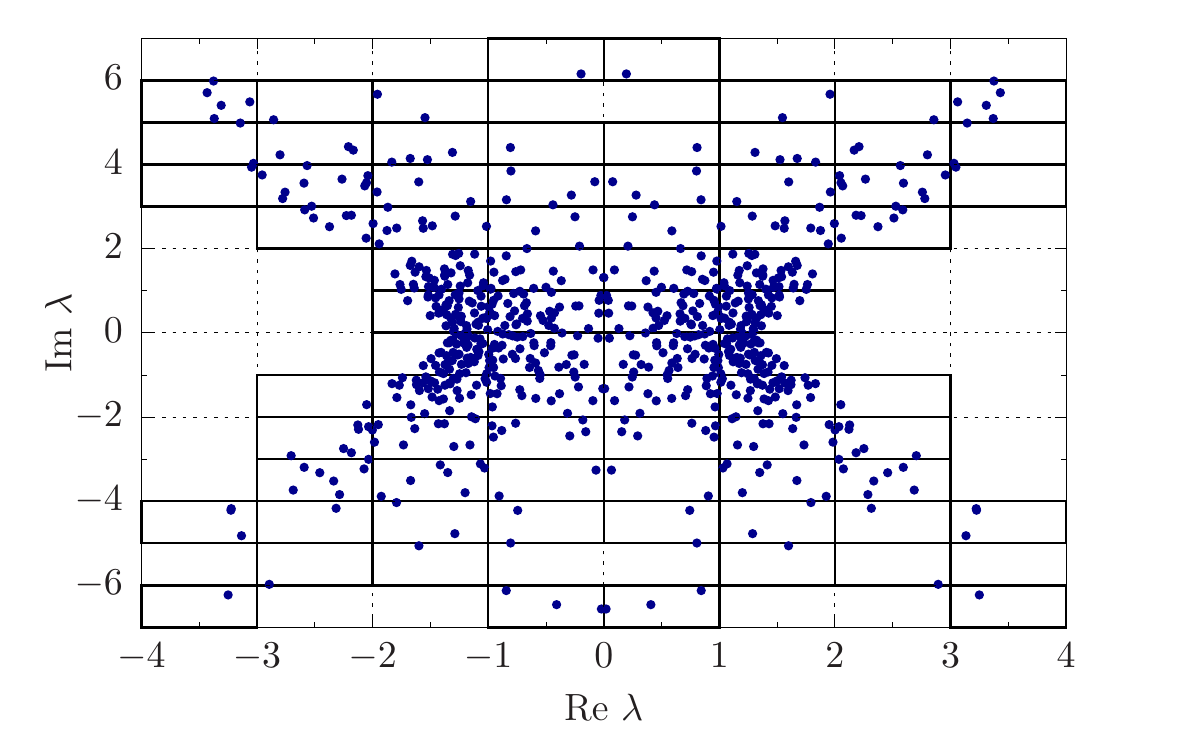}
\end{subfigure}%
\begin{subfigure}{.5\textwidth}
  \centering
  \includegraphics[width=.98\linewidth]{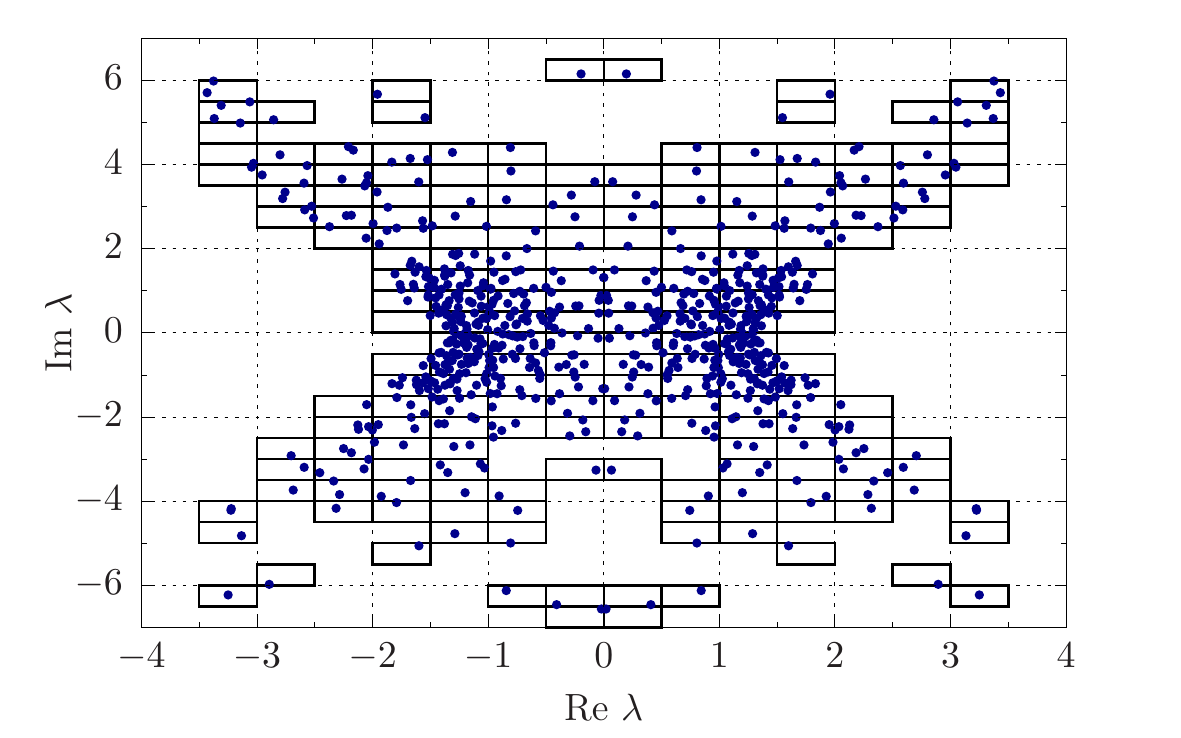}
\end{subfigure}
\begin{subfigure}{.5\textwidth}
  \centering
  \includegraphics[width=.98\linewidth]{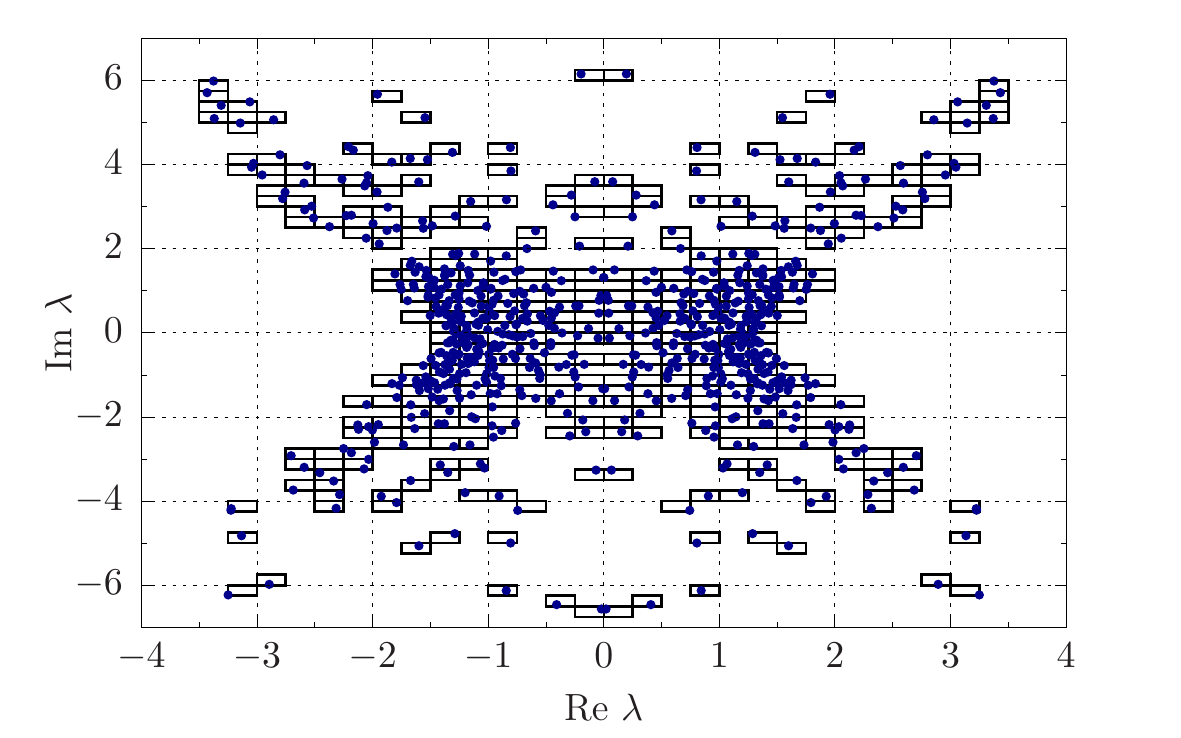}
\end{subfigure}%
\begin{subfigure}{.5\textwidth}
  \centering
  \includegraphics[width=.98\linewidth]{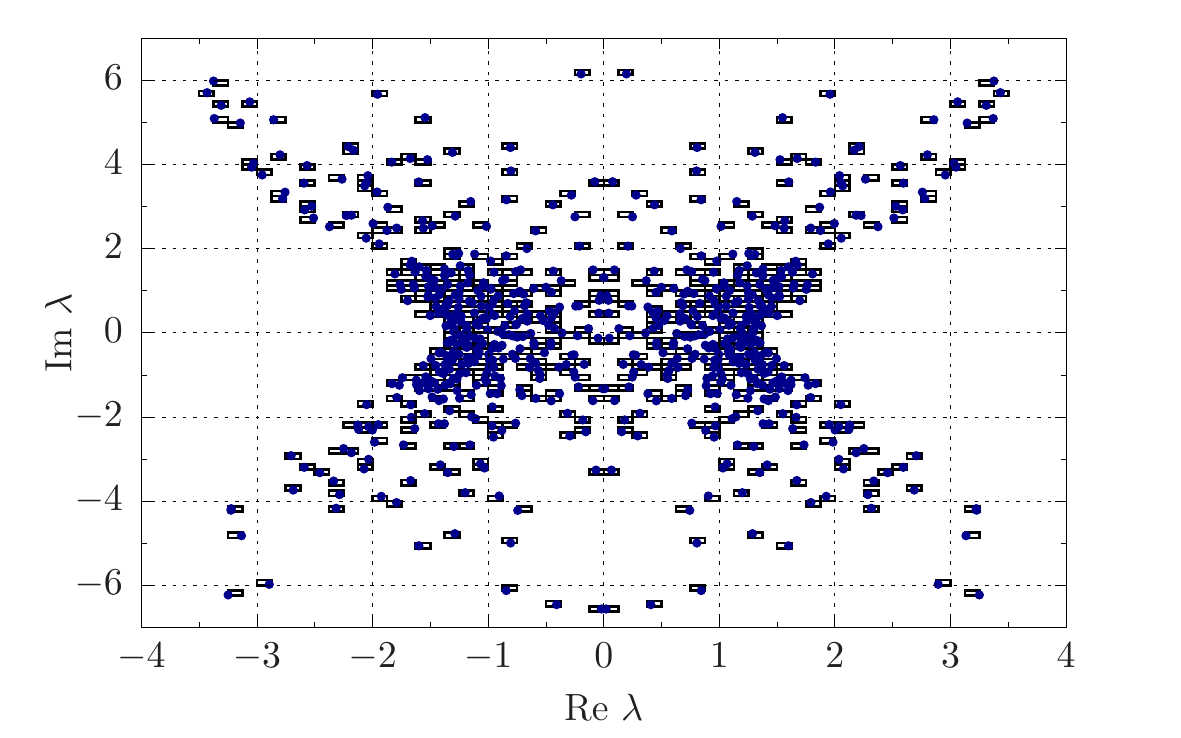}
\end{subfigure}
\caption{Illustration of the refinement.} 
\label{fig.refi.sketch}
\end{figure} 
Figure~\ref{fig.refi.sketch} is an illustration of this procedure, showing 4 steps in the refinement. The relative accuracy which can be achieved is shown in figure~\ref{fig.refi.acc}. The left panel shows the accuracy as a function of the size of the squares. The right panel shows the accuracy as a function of the order $k$.
\begin{figure}
\centering
\begin{subfigure}{.5\textwidth}
  \centering
  \includegraphics[width=1.0\linewidth]{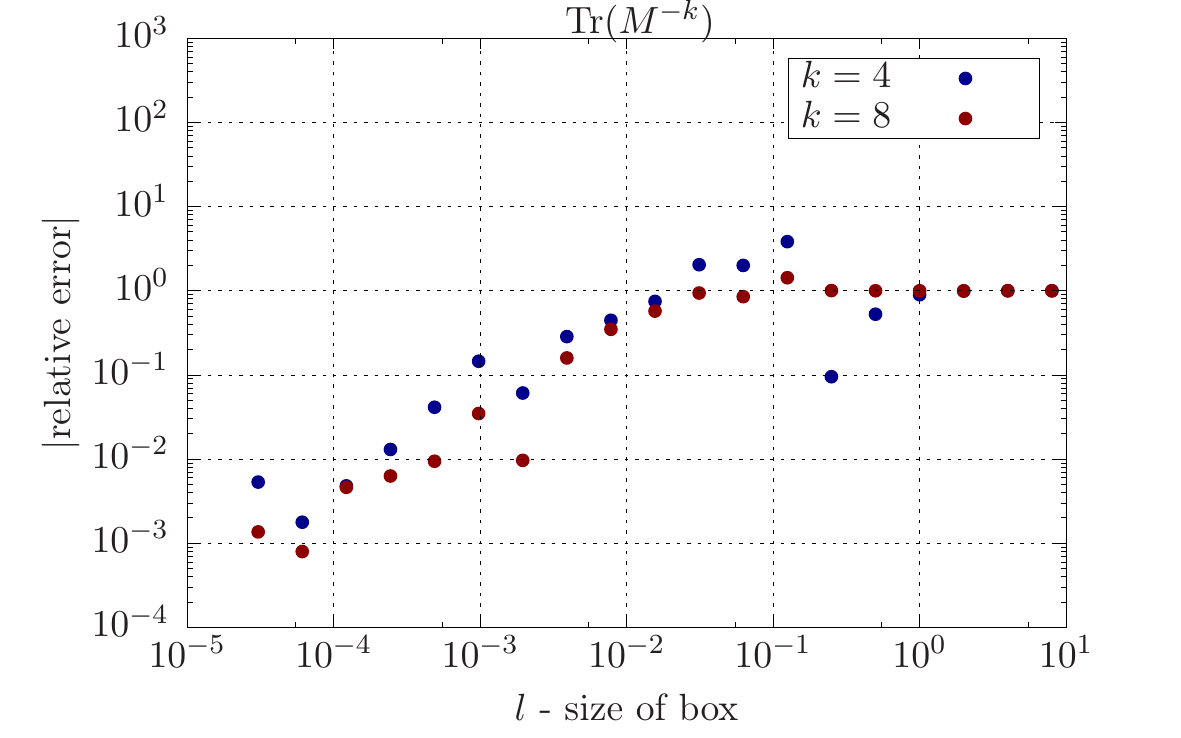}
\end{subfigure}%
\begin{subfigure}{.5\textwidth}
  \centering
  \includegraphics[width=1.0\linewidth]{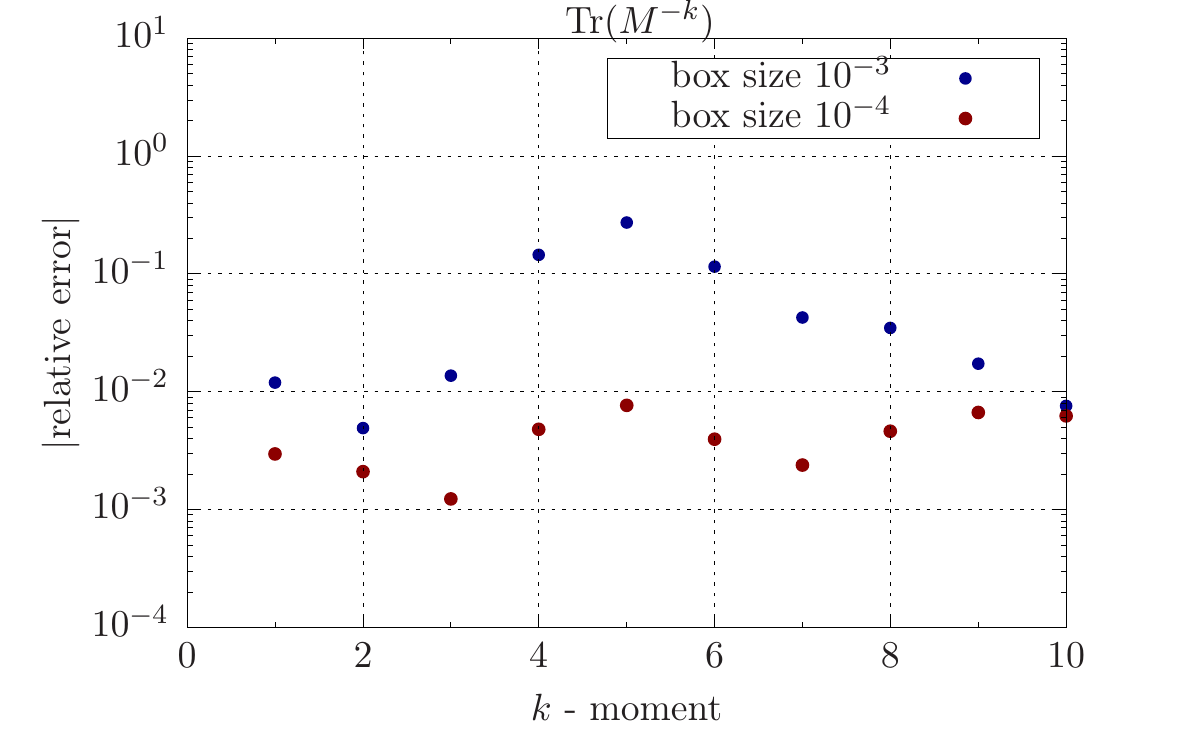}
\end{subfigure}
\caption{Relative accuracy of the refinement for different square-sizes (left) and moments (right).}
\label{fig.refi.acc}
\end{figure}
We note that the accuracy shown in figure~\ref{fig.refi.acc} is obtained using an exact estimate for the trace and contour integration in equation~(\ref{eq.refi.eig}). However, the number of squares will, especially in the final steps of the refinement, be close to the total number of eigenvalues, which requires a prohibitively large number of operations. A similar approach, in polar coordinates, has been used under the name of "cut Baum-Kuchen" algorithm in \cite{Nakamura:2013ska,Wakayama:2018wkc}.

\section{Single Contour Integration}
 
Instead of using a lot of squares to cover the complex plane, a single contour can be used to determine the relevant trace in equation~(\ref{eq.intro.tr}). Including a pole of order $k$ in the contour integral\footnote{We thank Tobias Rindlisbacher for suggesting this.} allows us to obtain the relevant trace, i.e.    
\begin{equation} 
\mathrm{Tr} \left[ \left( \slashed{D}^{-1} \frac{\partial \slashed{D}}{\partial \mu}\right)^k \right] \approx \frac{1}{n_I} \sum \limits_{j=1}^{n_I} z_j^{-k+1} \,
\mathrm{Tr} \left[ \left( z_j \mathds{1} - M \right)^{-1}\right] \label{eq.single.contour}
\end{equation}
where $z_j = r \mathrm{e}^{2 \pi \mathrm{i} j / n_I}$.
The circular contour on the right hand side of equation~(\ref{eq.single.contour}) must include no eigenvalues, so that the only pole is located at the origin. As before we study the inverse problem, i.e. we use 
\begin{equation}
M =  \slashed{D}\,\, \left( \frac{\partial \slashed{D}}{\partial \mu}\right)^{-1}.
\end{equation}
The trace is estimated using $n_V$ noise vectors. For the discretized circular contour we use $n_I$ equally spaced quadrature points. 
\begin{figure}
\centering
\begin{subfigure}{.5\textwidth}
  \centering
  \includegraphics[width=.98\linewidth]{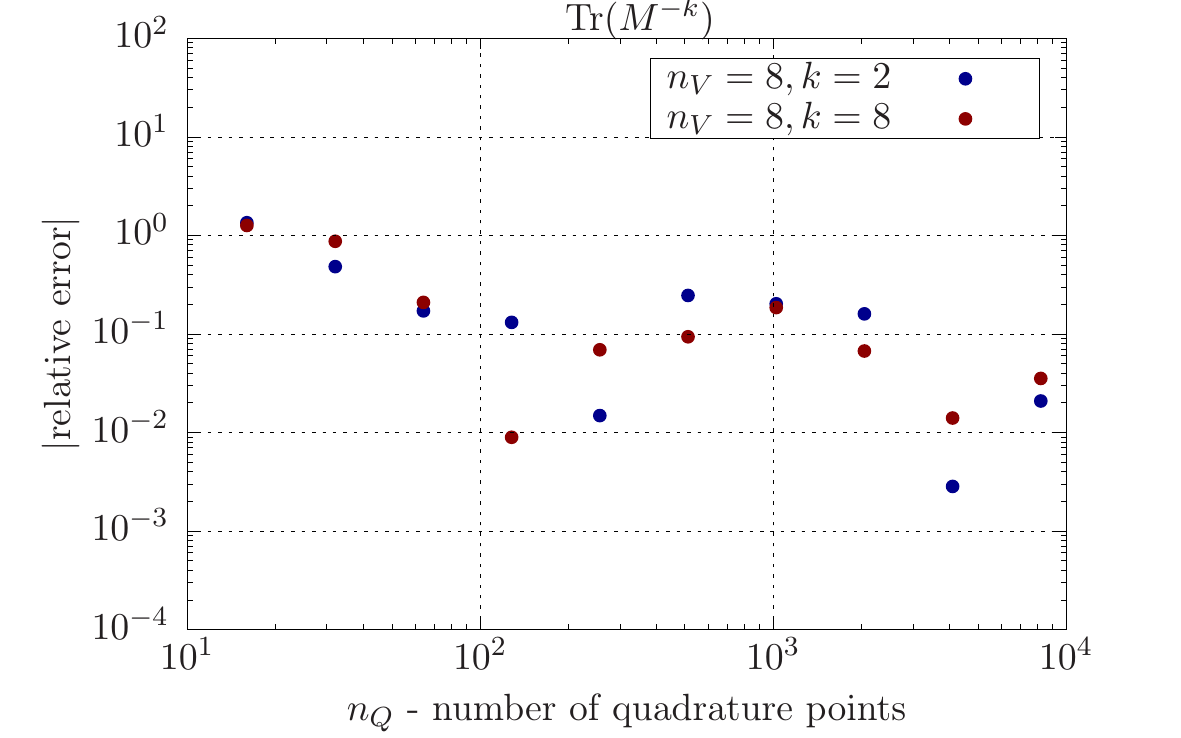}
\end{subfigure}%
\begin{subfigure}{.5\textwidth}
  \centering
  \includegraphics[width=.98\linewidth]{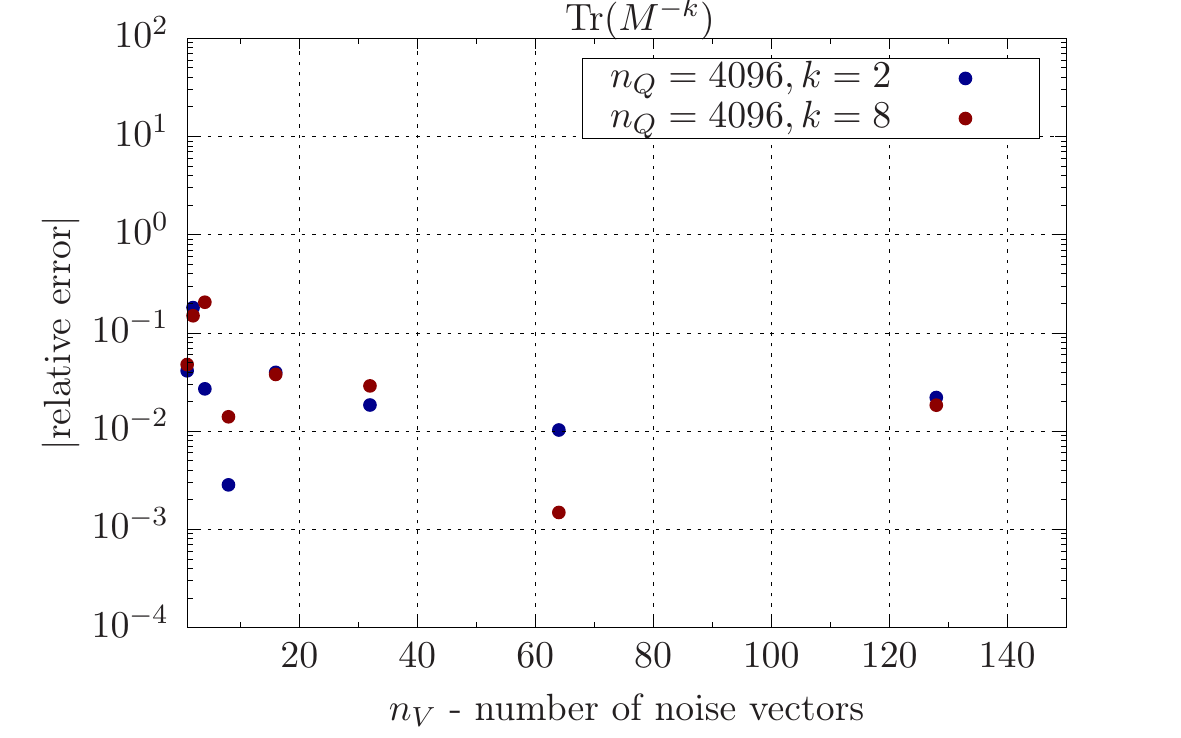}
\end{subfigure}
\begin{subfigure}{.5\textwidth}
  \centering
  \includegraphics[width=.98\linewidth]{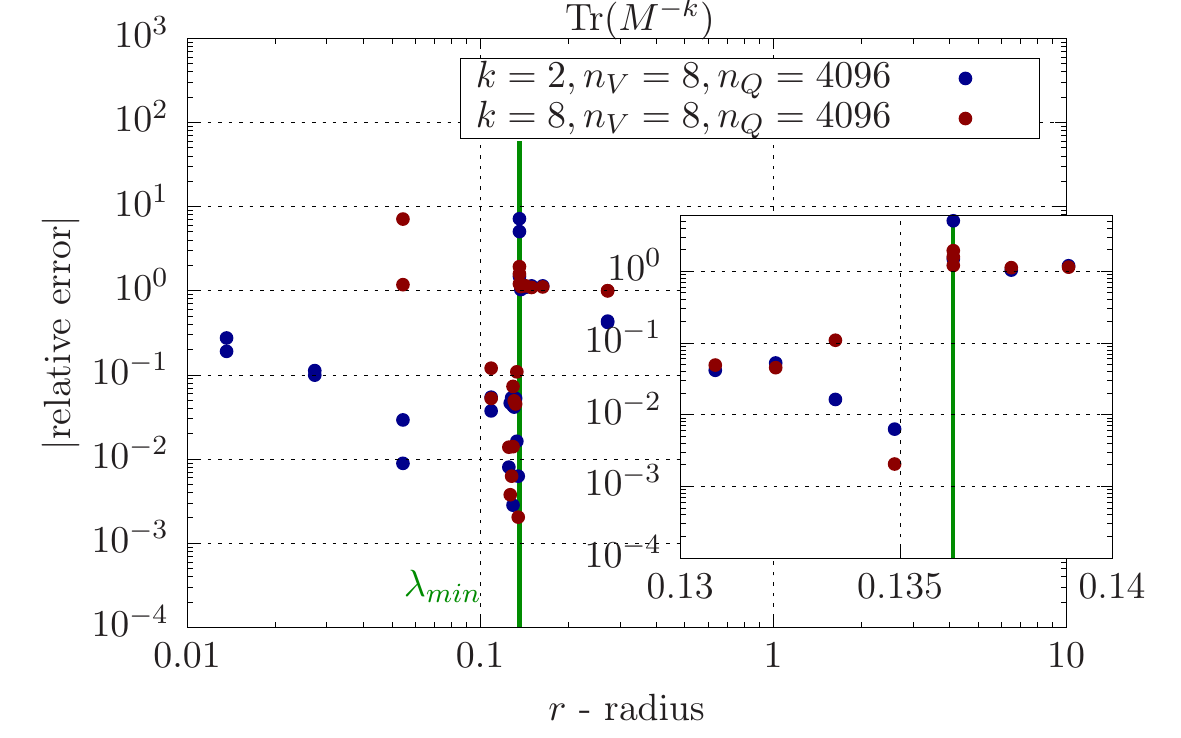}
\end{subfigure}%
\begin{subfigure}{.5\textwidth}
  \centering
  \includegraphics[width=.98\linewidth]{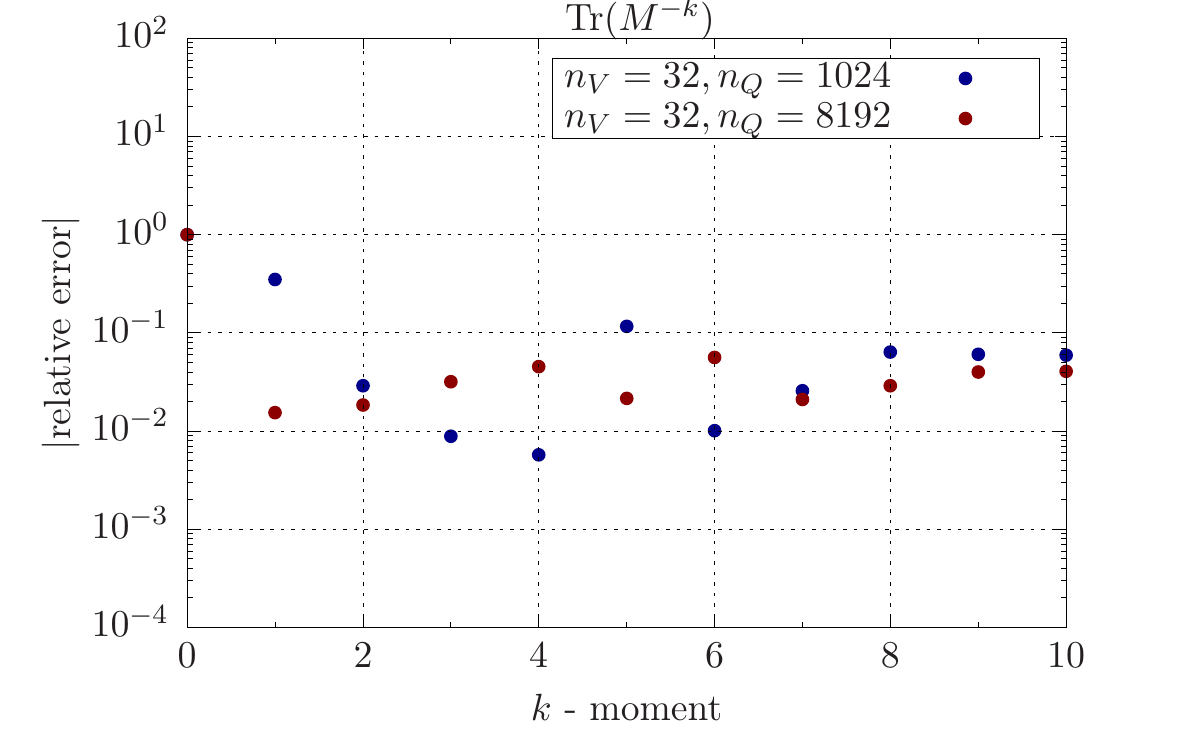}
\end{subfigure}
\caption{Relative accuracy using a single circular contour as a function of the number of integration points $n_I$ (top left), noise vectors $n_V$ (top right), radius of  contour $r$ (bottom left) and moment $k$ (bottom right).}
\label{fig.cont.para}
\end{figure}
Figure~\ref{fig.cont.para} shows the relative accuracy as a function of the number of integration points $n_I$ (top left), noise vectors $n_V$ (top right), radius $r$ of contour (bottom left) and moment $k$ (bottom right). As before, a shifted solver, especially a block solver as proposed in~\cite{deForcrand:2018orx}, can in principle determine the traces $\mathrm{Tr} \left[ \left( z_j \mathds{1} - A \right)^{-1}\right]$ efficiently. These traces are independent of the order $k$, so no additional solves are necessary for higher orders $k$. We find that the optimal choice for the circular contour is very close to the smallest eigenvalue (see fig. 4 bottom left, inset), which has to be determined before equation~(\ref{eq.single.contour}) can be exploited. Here we use a radius that is $95\%$ of the smallest eigenvalue, which can be obtained efficiently using for instance the Arnoldi method as implemented in ARPACK~\cite{lehoucq1998arpack,Arpack}.

\section{Summary}

We have presented an update on our efforts~\cite{deForcrand:2017cja} to determine the Taylor coefficients for finite-density QCD based on the Cauchy Residue Theorem. We have illustrated two distinct alternatives, where the first one relies on a refinement procedure to locate the eigenvalues of the relevant operator to sufficient precision. However, the numerical effort is too large to provide a practical alternative. On the other hand, the second approach uses a single discretized circular contour around the origin, which allows us to determine the relevant trace to an accuracy of a few percent. The recent development of shifted block solvers, as in~\cite{deForcrand:2018orx}, could make this approach practical, and perhaps competitive. Further improvements include a truncated solver~\cite{Bali:2009hu} or all-mode averaging~\cite{Blum:2012uh} and an improved integration scheme along the circular contour.

\section*{Acknowledgments}
 We thank Tobias Rindlisbacher for valuable discussion during early stages of this project. BJ acknowledges support by the Schweizerischer Nationalfonds (SNF) under grant 200020-162515 during his time at ETH Zürich.

\end{document}